\documentclass[acmsmall,nonacm,letterpaper]{acmart}

\usepackage{beramono}
\usepackage{calc}
\usepackage{listings}

\usepackage{acro}
\usepackage{mathpartir}
\usepackage{MnSymbol}

\AtBeginDocument{%
  }

\setcopyright{licensedusgovmixed}
\copyrightyear{2024}
\acmYear{2024}
\acmDOI{10.1145/3689492.3689814}

\acmConference[Onward! '24]{Proceedings of the 2024 ACM SIGPLAN International Symposium on New Ideas, New Paradigms, and Reflections on Programming and Software}{October 23--25, 2024}{Pasadena, CA, USA}

\acmBooktitle{Proceedings of the 2024 ACM SIGPLAN International Symposium on New Ideas, New Paradigms, and Reflections on Programming and Software (Onward! '24), October 23--25, 2024, Pasadena, CA, USA}
\acmISBN{979-8-4007-1215-9/24/10}

\DeclareAcronym{adt}{
  short = ADT,
  long  = abstract data type,
}

\makeatletter
\newlength{\linenumwidth} 
\newlength{\numwidth}%
\def\lst@PlaceNumber{%
  \makebox[\numwidth+1em][l]{%
    \makebox[\numwidth][r]{\normalfont\lst@numberstyle{\thelstnumber}}%
  }%
}
\makeatother

\newcommand*{\cochis}{\TirName{Cochis}}
\newcommand*{\fn}[1]{\ensuremath{\text{\itshape #1}}}
\newcommand*{\tuple}[1]{\ensuremath{\langle{#1}\rangle}}

\begin{document}

\title{System $F^\omega$ with Coherent Implicit Resolution}


\author{Eugène Flesselle}
\authornote{Category: graduate; ACM member number: 6914199; Research advisor: Dimi Racordon, EPFL}
\affiliation{%
  \institution{EPFL}
  \country{Switzerland}
}
\email{eugene.flesselle@epfl.ch}
\orcid{0009-0003-7545-594X}


\begin{abstract}
We propose a calculus for modeling implicit programming that supports first-class, overlapping, locally scoped, and higher-order instances with higher-kinded types.
We propose a straightforward generalization of the well-established System $F^\omega$ to implicit parameters, with a uniform treatment of type and term abstractions.
Unlike previous works, we give a \emph{declarative} specification of unambiguous, and thus coherent, resolution without introducing restrictions motivated by an algorithmic formulation of resolution.
\end{abstract}



\maketitle

\section{Introduction}

The extensibility of data abstractions is a well-known challenge for the design of programming languages
as first observed by Reynolds~\cite{Reynolds:TSLD1994} and coined as the \emph{expression problem} by Wadler~\cite{wadler:expression-problem}.
Type classes~\cite{DBLP:conf/popl/WadlerB89} have become a popular method to address this problem.
A type class is a set of operations that a type must support.
An implementation of these operations for a specific type forms a \emph{model} witnessing its conformance to the type class.
One can abstract over models, thereby lifting type-specific operations as constraints on the \emph{context} where abstractions can be applied.

The ergonomics of type classes hinge on the compiler's ability to supply models implicitly from type information,
conversely to the common ability to infer type arguments from term information.
These models are drawn from a set of values declared as available for implicit resolution, which we will refer to as \emph{given instances}.
More generally, implicit arguments can be constructed from instantiations and applications,
in the presence of polymorphic or parametric given instances, 
through a type-directed process known as \emph{implicit resolution}.
For example, the following Scala definition 
\lstset{keepspaces=true}
\lstinline[language=scala, morekeywords={given, using}]{given [A](using Show[A]): Show[List[A]] = ???}
provides an instance of \lstinline|Show[List[A]]| for any type \lstinline|A| in a context with a witness of \lstinline|Show[A]|.

A complication of type class oriented programming is that multiple given instances may derive distinct models of the same type.
Leading to potentially ambiguous choices during resolution, with semantically relevant consequences.
It is strongly desirable --- and even necessary in the presence of associated types~\cite{DBLP:conf/popl/ChakravartyKJM05, DBLP:conf/icfp/ChakravartyKJ05} --- to guarantee \emph{coherence}, the property that well-typed programs must have only one meaning~\cite{DBLP:conf/tacs/Reynolds91}.
There exists an ongoing debate on exactly which properties languages should uphold and how to do so, as we discuss next.

\vskip 0.05in
Languages subscribing to the Haskell school prefer resolution to be independent of the context in which it takes place. 
Models are second-class and restrictions are put in place to guarantee 
\emph{global uniqueness of type class instances}~\cite{url/Yang/uniqueness}, which imposes that any type have at most one given instance.
This approach is sufficient to ensure coherence but is severely anti-modular as defining an instance in one software component may prevent interoperability \cite{DBLP:conf/pldi/SiekL05, DBLP:conf/popl/DreyerHCK07, DBLP:conf/haskell/WinantD18}.
Moreover, the details of the numerous restrictions put in place are not only necessary for the language specification,
but also for users to reason about implicit resolution.
Finally, attempts to relax the aforementioned limitations often result in further convoluted systems --- Haskell's twelve type class related extensions~\cite{Gozdiovas24}, or Rust's orphan rules~\cite{rust/reference/orphan} --- and may even open soundness holes~\cite{DBLP:conf/popl/WeirichVJZ11}.

Conversely, proponents of dependent typing, 
such as Scala~\cite{DBLP:journals/pacmpl/OderskyBLBMS18} and Coq~\cite{DBLP:conf/tphol/SozeauO08},
typically advocate for models being first-class objects
governed by the same rules as other values, i.e. without scoping or overlapping restrictions.
Implicit resolution is then merely a way to omit certain arguments and immediately benefits the full flexibility and soundness properties from using a powerful dependently-typed language.
While these systems are far more permissive in the allowed derivations, they typically involve intricate disambiguation policies and lack completeness guarantees for the inference of these derivations.

While these approaches have been adopted by well-established languages and are generally assumed coherent, 
formal presentations are limited and rarely concise.
We recently proposed --- currently under review --- an extension of System $F$ with coherent and complete implicit resolution, $F_{CCI}$, built as an improvement on \cochis{}~\cite{DBLP:journals/jfp/SchrijversOWM19}, which already supported first-class, locally scoped, overlapping, and higher-order given instances while remaining coherent.
Like \cochis{}, we first give a concise but ambiguous judgment for implicit resolution.
Unlike \cochis{}, who then introduces a number of restrictions to obtain a deterministic --- and thus coherent --- algorithmic formulation,
we propose a declarative specification of unambiguous resolution, unconcerned with details of implementation.
Even though the problem boils down to one of type inhabitation~\cite{master:Rouvoet16}, which is undecidable~\cite{10.5555/1197021, DBLP:conf/types/DudenhefnerR18},
we provide an algorithmic --- but not necessarily terminating --- formulation of resolution that is complete w.r.t. the declarative specification.
In this abstract, we propose a further extension of the declarative system to higher-kinded types,
which is, to our knowledge the first formalization to do so.
Adapting the algorithmic version remains the central task of ongoing work.

\section{The $F^\omega_{CCI}$ Calculus}

\begin{mathpar}
\begin{tabular}{r c l l c c c r}
  \text{Kinds} & $\kappa$ & $::=$ & $\ast \mid \kappa\to\kappa$ 
            &&&&
            \\
  \text{Simple Types} & ${\sigma}$ & $::=$ &
            $\alpha$
            $\mid \Pi(\tau).\tau$
            $\mid \forall (\alpha : \kappa) . \tau$
            $\mid \alpha\ \tau$
            &&&&
            $\tau_1 \rightarrow \tau_2 := \Pi(\tau_1).\tau_2$ \\
  \text{Types} & ${\tau}$ & $::=$ &
            $\sigma$
            $\mid \Pi[\tau].\tau$
            $\mid \forall [\alpha : \kappa] . \tau$ 
            $\mid \lambda \alpha : \kappa . \tau \mid \tau\ \tau$
            &&&&
            $\tau_1\Rightarrow\tau_2 := \Pi[\tau_1].\tau_2$ \\
  \text{Terms} & $e$ & $::=$ &
            $x \mid$
            ${\lambda \tuple{x : \tau} . e} \mid$
            ${e \tuple{e}} \mid$
            ${\Lambda \tuple{\alpha : \kappa} . e} \mid$
            ${e \tuple{\tau}} \mid$
            ${e :: \tau}$ 
            &&&& 
            $\tuple{\cdot} ::= ( \cdot ) \mid [ \cdot ]$
            \\
  \text{Contexts} & ${\Gamma}$ & $::=$ &
            $\varepsilon \mid$
            $\Gamma, \tuple{\alpha : \kappa} \mid$
            $\Gamma, \tuple{x : \tau}$
            &&&&
            $\tuple{\alpha} := \tuple{\alpha : \ast}$ \\
\end{tabular}
\end{mathpar}

Our syntax extends System $F^\omega$ with support for contextual abstractions: functions taking parameters passed implicitly, which we will refer to as \emph{implicit parameters}.
We allow both term and type abstractions to specify whether their parameters are taken explicitly or implicitly, via the use of the $(\cdot)$ and $[\cdot]$ binder forms respectively.
For example, $\lambda (x: \tau) . x$ denotes a usual function whereas $\lambda [x: \tau] . x$ is a context function whose argument, which we will call a \emph{requirement}, can be passed implicitly.
Accordingly, types $\tau$ extend those of system $F^\omega$ with implicit function types $\tau_1 \Rightarrow \tau_2$ and implicit universal types $\forall[\alpha : \kappa].\tau$.
We admit the trivial extensions to $F^\omega_{CCI}$ of the kinding $\Gamma \vdash \tau : \kappa$ and type equivalence $\Gamma \vdash \tau \equiv \tau'$ judgments.
\begin{mathpar}
  \inferrule[t-var]
    {\langle x : \tau \rangle \in \Gamma }
    {\Gamma \vdash x : \tau}
  \and
  \inferrule[t-abs]
    {\Gamma \vdash \tau_1 : \ast \quad\ \Gamma, \tuple{x : \tau_1} \vdash e : \tau_2}
    {\Gamma \vdash \lambda \tuple{x : \tau_1} . e : \Pi \tuple{\tau_1} . \tau_2}
  \and
  \inferrule[t-app]
    {\Gamma \vdash e_1 : \Pi \tuple{\tau_2} . \tau_1 \quad\
      \Gamma \vdash e_2 : \tau_2}
    {\Gamma \vdash e_1 \langle e_2 \rangle : \tau_1}
  \\
  \inferrule[t-tabs]
    {\Gamma, \tuple{\alpha : \kappa} \vdash e : \tau}
    {\Gamma \vdash \Lambda \tuple{\alpha : \kappa} . e : \forall \tuple{\alpha : \kappa} . \tau}
  \and
  \inferrule[t-tapp]
    {\Gamma \vdash e : \forall \tuple{\alpha : \kappa} . \tau_1 \quad\
      \Gamma \vdash \tau : \kappa}
    {\Gamma \vdash e \langle \tau \rangle : [\alpha \mapsto \tau]\tau_1}
  \and
  \inferrule[t-asc]
    {\Gamma \vdash e : \tau_1 \quad
      \Gamma \Vdash \tau_1 \leadsto \tau \quad
      \Gamma \vdash \tau : \ast
    }
    {\Gamma \vdash e :: \tau : \tau}
\end{mathpar}

The first five typing rules are a trivial extension of $F^\omega$, where the only change is the generalization of binder constructs.%
\footnote{We assume that it is understood all occurrences of $\tuple{\cdot}$ in a rule are the same.}
Context functions $\lambda [x : \tau]. e$ allow expressing a \emph{requirement} from the contexts in which they can be applied, which is recorded in their type to later allow taking their arguments implicitly.%
\footnote{One novel aspect of $F^\omega_{CCI}$ is that we handle type abstractions in a perfectly symmetric way.}
In addition, their body is type-checked in a context extended with the assumption $[x : \tau]$ denoting a binding available to implicit resolution.
Symmetrically, explicit applications of context functions $e_1 [e_2]$ are a way to \emph{provide} a concrete value to the implicit context.\footnote{Though not necessary for soundness, we require applications to use a binder form consistent with the type of the abstraction, to avoid accidental applications of context functions.}
The corresponding derived form $\fn{let}\ [x : \tau] = e_1\ \fn{in}\ e_2 := (\lambda [x : \tau]. e_2) [e_1]$ is equivalent to Scala's \lstinline|given| construct.

\emph{Querying} the implicit context is done with ascriptions $e :: \tau$ which trigger the implicit resolution judgment to fulfill the requirements of $e$.
An alternative design~\cite{DBLP:journals/jfp/SchrijversOWM19, DBLP:conf/pldi/OliveiraSCLY12} is to have a built-in query operator $?\tau$.
Contrarily to our system, such an approach does not simultaneously cover the inference of type parameters 
or support the resolution of multiple implicit arguments at once.
In fact, the query operator can be expressed in $F^\omega_{CCI}$ with the context abstraction
$\fn{summon} := \Lambda[\alpha].\lambda[x : \alpha].x$ of type $\forall[\alpha].\alpha\Rightarrow\alpha$,
used for example as follows $\fn{let}\ [y : \fn{Int}] = 7\ \fn{in}\ \fn{summon} :: \fn{Int}$.

Finally, remark that implicit resolution is \emph{only} triggered at explicit ascriptions, similarly to the notion of \emph{non-maximally inserted} implicits in Coq~\cite{DBLP:conf/tphol/SozeauO08}, as opposed to allowing \TirName{t-asc} to apply anywhere, similarly to a subsumption rule.
We opt for explicit ascriptions however since, unlike subtyping, implicit resolution is not semantically irrelevant.
Furthermore, the alternative approach would entail losing the syntax-directedness of the typing rules,
and the principal type property.%
\footnote{There is indeed no principal choice between keeping a contextual abstraction unspecialized or capturing local instances.}
\begin{mathpar}
  \inferrule[m-beta]
    {\Gamma \Vdash [\alpha \mapsto \tau_2] \tau_1 \leadsto \sigma}
    {\Gamma \Vdash (\lambda \alpha : \kappa . \tau_1) \tau_2 \leadsto \sigma}
  \and
  \inferrule[m-all]
    {\Gamma \vdash \tau : \kappa \quad\ \Gamma \Vdash [\alpha\mapsto\tau]\tau_1 \leadsto \sigma}
    {\Gamma \Vdash \forall[\alpha : \kappa].\tau_1 \leadsto \sigma}
  \and
  \inferrule[m-impl]
    {\Gamma \Vdash \tau_1 \quad\ \Gamma \Vdash \tau_2 \leadsto \sigma}
    {\Gamma \Vdash \tau_1\Rightarrow\tau_2 \leadsto \sigma}
  \and
  \inferrule[m-equiv]
    {\Gamma \vdash \sigma_1 \equiv \sigma}
    {\Gamma \Vdash \sigma_1 \leadsto \sigma}
  \\
  \inferrule[f-beta]
    {\Gamma \Vdash \tau \leadsto [\alpha \mapsto \tau_2] \tau_1}
    {\Gamma \Vdash \tau \leadsto (\lambda \alpha : \kappa . \tau_1) \tau_2}
  \and
  \inferrule[f-all]
    {\Gamma,[\alpha : \kappa] \Vdash \tau \leadsto \tau_1}
    {\Gamma \Vdash \tau \leadsto \forall[\alpha : \kappa].\tau_1}
  \and
  \inferrule[f-impl]
    {\Gamma,[x : \tau_1] \Vdash \tau \leadsto \tau_2}
    {\Gamma \Vdash \tau \leadsto \tau_1\Rightarrow\tau_2}
  \and
  \inferrule[lookup]
    {[x : \tau] \in \Gamma \quad\ \Gamma \Vdash \tau \leadsto \tau_1}
    {\Gamma \Vdash \tau_1}
\end{mathpar}

The declarative judgment of implicit resolution $\Gamma \Vdash \tau \leadsto \tau'$, read as $\Gamma$ entails that $\tau$ matches $\tau'$,
holds if the implicit parameters of a term of type $\tau$ can be derived from $\Gamma$ to obtain one of type $\tau'$.%
\footnote{Although omitted here in the interest of space, this judgment typically also gives the elaboration to terms of System~$F^\omega$.}

If $\tau'$ is a simple type $\sigma$ then we proceed based on the structure of $\tau$.
If it is a type operator application, then we simply reduce it to continue resolution.
If $\tau$ is an implicit universal type $\forall[\alpha:\kappa].\tau_1$, then it can be instantiated with any appropriately kinded type to continue matching $\sigma$.
If $\tau$ is an implicit function type $\tau_1\Rightarrow\tau_2$, then we need $\tau_2$ to match $\sigma$ and for $\Gamma$ to derive $\tau_1$, which holds if the latter is matched by any type with an implicit binding $[x : \tau] \in \Gamma$.
Finally, $\sigma$ is matched by any other simple type equivalent to it.
Otherwise, if $\tau'$ not a simple type, then the entailment relation introduces all abstraction parameters into the context, in addition to reducing type operator applications.
This approach, known as \emph{focusing} in proof search~\cite{DBLP:journals/logcom/Andreoli92}, was already adapted by \cochis{}.
The idea is to support typings like
$[y : \tau_2], (f : \tau_1\Rightarrow\tau_2\Rightarrow\tau_3) \vdash f :: \tau_1\Rightarrow\tau_3$
by performing resolution of contextual abstractions in environments assuming their requirements.

One can easily notice the choice of $\tau$ is ambiguous in both \TirName{lookup} and \TirName{m-all}.
Many refinements can be made to eliminate these sources of ambiguities, but not without rejecting desirable programs.
Instead, we simply define unambiguous resolution by taking the derivation of the \emph{least} depth\footnote{Which can be more formally defined on witnesses, once the elaboration semantics have been presented.}.
In the case of multiple \emph{minimal} instances, an ambiguity error can be reported.

\vskip 0.1in
We have shown it is relatively straightforward to define --- and has been mechanized in Coq --- a \emph{coherent declarative resolution judgment} supporting higher-kinded types without introducing syntactic restrictions.
Subsequently, much of the complexity lies in providing an algorithmic formulation guaranteeing completeness, which has been the focus of our ongoing research.

\bibliographystyle{ACM-Reference-Format}
\bibliography{references}

\end{document}
\endinput